\documentclass[11pt,twoside]{article}
\usepackage{epsfig}
\usepackage{here}
\usepackage{amsmath}

\begin{document}
{\hfill RUB-TPII-05/04}\\ [1cm]
\begin{center}{\bf\Large The gluon content of the $\mathbf{\eta}$ and
       $\mathbf{\eta^{\prime}}$ mesons\footnote{Invited
        plenary talk presented by the first author at \textit{Second
        Symposium on Threshold Meson Production in pp and pd
        Interactions}. Extended COSY-11 collaboration meeting. Collegium
        Maius, Cracow, Poland, May 31 - June 3, 2004.}}
\end{center}
\vspace{0.5cm}
\begin{center}
   N.~G.~Stefanis\footnote{Corresponding author:
   stefanis@tp2.ruhr-uni-bochum.de}$^a$ %
   and S.~S.~Agaev$^b$ %
   \\[0.1cm]
   {\small \em
         $^a$Institut f\"ur Theoretische Physik II, Ruhr-Universit\"at
         Bochum, D-44780 Bochum, Germany \\
         $^b$High Energy Physics Lab., Baku State University,
         Z.\ Khalilov St.\ 23, 370148 Baku, Azerbaijan
   }
\end{center}
\vspace{0.5cm}
\begin{center}
 \parbox{0.9\textwidth}{
  \small{
    {\bf Abstract:}\
     We analyze power-suppressed contributions to singlet pseudoscalar
     $\eta$ and $\eta^{\prime}$ meson transition form factors.
     These corrections stem from endpoint singularities and help
     improve the agreement between QCD theory and the experimental data,
     in particular, at low-momentum transfers. Using the CLEO data,
     we extract information on the profile of the $\eta_1$ and $\eta_8$
     distribution amplitudes in the $SU(3)_{F}$ octet-singlet basis
     employing both the one-angle and the two-angle mixing schemes.
     In the former scheme, we find good agreement with the CLEO data,
     while in the second case, our approach requires
     non-asymptotic profiles for these mesons.
     }
 }
\end{center}

\vspace{0.5cm}
\section{Introduction}
\label{sec:intro} Control over power-behaved corrections in QCD
processes is crucial for the correct interpretation of
high-precision experiments in which intact hadrons appear in the
initial and/or final states. Prominent examples are meson-photon
transition form factors, as measured by the CLEO collaboration
\cite{CLEO98} for the pion and the $\eta$ and $\eta^{\prime}$, and
the recent JLab high-precision data for the pion's electromagnetic
form factor \cite{JLab00}.

Because, theoretically, the dynamics of such exclusive processes
involves the corresponding meson distribution amplitudes, one can
extract crucial information about the nonperturbative partonic
structure of pseudoscalar mesons. In contrast to the
hard-scattering amplitude, that can be systematically computed
within perturbative QCD and is specific for each process, hadron
distribution amplitudes are universal quantities that encode the
partonic structure of hadrons. Their computation requires the
application of nonperturbative methods, like QCD sum rules with
nonlocal condensates---introduced in \cite{MR86,BR91,MS93} and
recently improved in \cite{BMS01}---to derive a realistic pion
distribution amplitude complying with the CLEO data on the
pion-photon transition at the $1\sigma$ level \cite{BMS02}. In
addition, this type of pion distribution amplitude was recently
\cite{BPSS04} used in conjunction with fixed-order
\cite{SS97,Shi-an} and resummed \cite{SSK99,KS01} Analytic
Perturbation Theory to calculate the pion's electromagnetic form
factor providing very good agreement with the existing data.

Alternatively, one can use the factorization QCD approach in order
to extract the shape of the pion (pseudoscalar meson) distribution
amplitude directly from the data. The reliability of the latter
possibility depends, however, on the way one deals with
fixed-order perturbative calculations. To minimize the influence
of (disregarded) higher-order contributions, while approaching the
kinematic endpoint regions of the process in question (where the
nonperturbative dynamics dominate), one may use Borel resummation
techniques and, by this way, estimate power-behaved corrections.
Indeed, this type of approach \cite{Ag04} was recently used to
compute the pion-photon transition form factor and determine the
pion distribution amplitude with results for the latter quantity
close to the profiles determined with the nonlocal QCD sum rules
just mentioned.

In the present exposition, we apply this type of approach (called
the RC method) to extract the distribution amplitudes of the
$\eta$ and $\eta^{\prime}$ mesons with particular focus being
placed on their gluonic content \cite{AS03}. More details of our
approach can be found in \cite{Aga01,AM01,AS02,Aga95a,Aga95b}.

\section{$\eta$-$\eta^{\prime}$ mixing schemes}
\label{sec:mix-schemes}
Physical pseudoscalar mesons $\eta$,
$\eta^\prime$ are admixtures of $SU(3)_{\rm F}$ octet ($\eta_8$)
and singlet ($\eta_1$) states:
\begin{equation}
\left(\begin{array}{c}\eta \\
                      \eta^{\prime}
      \end{array}\right)
=
\left(\begin{array}{cc}\cos{\theta_p} & -\sin{\theta_p}\\
                       \sin{\theta_p} &  \cos{\theta_p}\\
      \end{array}\right)
\left(\begin{array}{c}\eta_8 \\
                      \eta_1 \\
      \end{array}\right)\, ,
\label{eq:octet-singlet}
\end{equation}
where $\theta_p$ is the pseudoscalar mixing angle in the
octet-singlet scheme (for a review and further references, see
\cite{Fel99}).

On the parton level, the states $\eta_8$ and $\eta_1$ are given by
\begin{equation}
\left(\begin{array}{c}\eta_8 \\
                      \eta_1
      \end{array}\right)
=
\left(\begin{array}{cc}\sin{\theta_I} & -\cos{\theta_I}\\
                       \cos{\theta_I} &  \sin{\theta_I}\\
      \end{array}\right)
\left(\begin{array}{c}\frac{1}{\sqrt{2}}
\left(u\bar{u}+d\bar{d}\right) \\
                      s\bar{s} \\
      \end{array}\right)\, ,
\label{eq:partons}
\end{equation}
with $\theta_I$ being the ideal mixing angle.

Then, in turn, $\eta$ and $\eta^{\prime}$ are admixtures of
$q\bar{q}$ pairs (in a quark-flavor basis) expressed via
\begin{equation}
\left(\begin{array}{c}\eta \\
                      \eta^{\prime}
      \end{array}\right)
=
\left(\begin{array}{cc}\cos{\alpha_p} & -\sin{\alpha_p}\\
                       \sin{\alpha_p} &  \cos{\alpha_p}\\
      \end{array}\right)
\left(\begin{array}{c}\frac{1}{\sqrt{2}}
\left(u\bar{u}+d\bar{d}\right) \\
                      s\bar{s} \\
      \end{array}\right)\, ,
\label{eq:quark-flavor}
\end{equation}
where $\alpha_p=\theta_p-\theta_I+\pi/2$ denotes the deviation of
the mixing angle from the ideal one due to the $U_{A}(1)$
anomaly---in contrast to the vector meson $\phi\,-\,\omega$ system
with $\alpha_v\simeq 0$. Note that the flavor-singlet pseudoscalar
state contains also a gluon component: ``gluonium''. To
accommodate the gluonic component, one has to extend the mixing
scheme to a $3\times 3$ matrix with three mixing angles;
i.e.,\cite{Kou01} {\footnotesize
\begin{eqnarray}
\left(\!\!\!\begin{array}{c}\eta \\
                      \eta^{\prime}\\
                      \iota
      \end{array}\!\!\!\right)
\!\!\!\! & = & \!\!\!\! \left(\!\!\!\begin{array}{ccc}
 \cos{\theta_p}\cos{\gamma}+\sin{\theta_p}\cos{\phi}\sin{\gamma} &
-\sin{\theta_p}\cos{\gamma}+\cos{\theta_p}\cos{\phi}\sin{\gamma} &
 \sin{\phi}\sin{\gamma} \\
 \cos{\theta_p}\sin{\gamma}+\sin{\theta_p}\cos{\phi}             &
 \sin{\theta_p}\sin{\gamma}+\cos{\theta_p}\cos{\phi}\cos{\gamma} &
 \sin{\phi}\cos{\gamma} \\
-\sin{\theta_p}\sin{\phi}                                        &
-\cos{\theta_p}\sin{\phi}                                        &
 \cos{\phi} \\
      \end{array}\!\!\!\right) \nonumber \\
&& \times \left(\!\!\!\begin{array}{c}\eta_8 \\
                      \eta_1 \\
                      G
      \end{array}\!\!\!\right)\, ,
\label{eq:gluonic}
\end{eqnarray}}
where $\iota$ is a Glueball state and $G=|gg \rangle$ denotes
gluonium. Note that $|\eta\rangle \simeq |\eta_8\rangle$ (because
$m_\eta \simeq m_8$ due to the Gell-Mann--Okubo mass formula), so
that the $|\eta_{1}\rangle$ admixture is small with practically no
room for a $|G\rangle$ contribution. Hence, $\gamma=0$, and, as a
result,
\begin{equation}
\left(\!\!\!\begin{array}{c}\eta \\
                      \eta^{\prime}\\
                      \iota
      \end{array}\!\!\!\right)
= \left(\!\!\!\begin{array}{ccc}
 \cos{\theta_p}    &
-\sin{\theta_p}    &
 0                 \\
 \sin{\theta_p}\cos{\phi} &
 \cos{\theta_p}\cos{\phi} &
 \sin{\phi}        \\
-\sin{\theta_p}\sin{\phi} &
-\cos{\theta_p}\sin{\phi} &
 \cos{\phi}         \\
\end{array}\!\!\!\right)
\left(\!\!\!\begin{array}{c}\eta_8 \\
                      \eta_1 \\
                      G
      \end{array}\!\!\!\right)\, .
\label{eq:noG}
\end{equation}
A physical state is then a superposition of the sort
\begin{equation}
  |\psi\rangle
=
  x|Q\rangle + y|S\rangle + z|G\rangle \, ,
  \qquad x^2 + y^2 + z^2 = 1
\label{eq:superposition}
\end{equation}
with components (in the quark-flavor basis) given by
\begin{eqnarray}
|Q\rangle &=& \frac{1}{\sqrt{2}} \left(u\bar{u}+d\bar{d}\right)
\nonumber \\
|S\rangle &=& |s\bar{s}\rangle \nonumber \\
|G\rangle &=& |gg\rangle \, .
\label{eq:components}
\end{eqnarray}

The physical states $\eta$ and $\eta^\prime$ are
\begin{equation}
|\eta\rangle = x_{\eta}|Q\rangle + y_{\eta}|S\rangle \, , \qquad
  |\eta^\prime \rangle = x_{\eta^\prime}|Q\rangle
  + y_{\eta^\prime}|S\rangle
  + z_{\eta^\prime}|G\rangle
\label{eq:physical}
\end{equation}
with mixing coefficients
\begin{equation}
 x_{\eta}^2 + y_{\eta}^2 = 1 \, ,
\qquad
  x_{\eta^\prime}^2 + y_{\eta^\prime}^2 + z_{\eta^\prime}^2 = 1
\label{eq:mixicoef}
\end{equation}
related to the mixing angles
\begin{equation}
x_{\eta} = \cos{\alpha_p}\, , \qquad y_{\eta}=-\sin{\alpha_p}
\label{eq:mixing-angle}
\end{equation}
and
\begin{equation}
x_{\eta^\prime}=\cos{\phi}\sin{\alpha_p}\, , \quad
y_{\eta^\prime}=\cos{\phi}\cos{\alpha_p}\, , \quad
z_{\eta^\prime}=\sin{\phi}\, .
\label{eq:mix-angle-prime}
\end{equation}
The $SU(3)_{\rm F}$ octet-singlet basis is provided by
\begin{eqnarray}
  |\eta_1\rangle
& \, = \, &
  \frac{1}{\sqrt{3}}\; |u\bar{u} + d\bar{d} + s\bar{s}\rangle
  \nonumber \\
  |\eta_8\rangle
& \, = \, &
  \frac{1}{\sqrt{6}}\; |u\bar{u} + d\bar{d} - 2s\bar{s}\rangle \, .
\label{eq:octet-singlet-basis}
\end{eqnarray}
One notes that the octet-singlet and the quark-flavor basis are
equivalent, but that the parameterizations of the decay constants
are different.

Let us now have a closer look to the decay constants of $M= \eta ,
\eta^\prime $ mesons. Their parameterization is defined via
\begin{equation}
 \langle 0| J_{\mu 5}^{i} |M\rangle = i f_{P}^{i} p_{\mu}\, ,
\label{eq:decay-const}
\end{equation}
where $J_{\mu 5}^{i}\,$ is the axial-vector current ($i=Q, S$ or
$i=1, 8$). In the quark-flavor basis, the decay constants follow
the pattern of state mixing, i.e.,
\begin{eqnarray}
  f_{\eta}^{Q}
& = &
  f_Q \cos{\alpha_p} \, \qquad f_{\eta}^{S} = -f_{S} \sin{\alpha_p}
  \nonumber \\[0.5cm]
f_{\eta^\prime}^{Q} & = &
  f_{Q}\sin{\alpha_p} \, \qquad f_{\eta^\prime}^{S} =
  f_{S}\cos{\alpha_p}\, .
\label{eq:decay-constquark-flavor}
\end{eqnarray}
In the octet-singlet basis the situation is different:
\begin{eqnarray}
  f_{\eta}^{8}
& = &
  f_8 \cos{\theta_8} \, \qquad f_{\eta}^{1} = -f_{1} \sin{\theta_1}
  \nonumber \\[0.5cm]
  f_{\eta^\prime}^{8}
& = &
  f_{8}\sin{\theta_8} \, \qquad f_{\eta^\prime}^{1} =
  f_{1}\cos{\theta_1}\, .
\label{eq:decay-const-octet-singlet}
\end{eqnarray}
Note that in general, $\theta_8\neq \theta_1 \neq \theta_p$. In
the present analysis, we use the octet-singlet basis with the
one-angle (standard) parameterization with
\begin{eqnarray}
& f_\pi = 0.131\mbox{GeV}\, \qquad f_{1} = 1.17 f_\pi  \notag \\[0.5cm]
& f_{8} = 1.26 f_\pi \, \qquad \theta_p = -15.4^\circ,
\label{eq:our-decay-const}
\end{eqnarray}
and the two-angle mixing scheme with the parameters
\begin{equation}  \label{eq:2angle}
\theta_p=-15.4^\circ,\;\theta_1=-9.2^\circ,\;
\theta_8=-21.2^\circ.
\end{equation}

\section{Electromagnetic $\eta\gamma$, $\eta^{\prime}\gamma$
         transition form factor}
\label{sec:elm-trn-ff}

In the Standard Hard-scattering Approach (HSA), the transverse
momenta are neglected (collinear approximation) and the meson
($\pi$, $\eta$, $\eta^{\prime}\ldots$) consists in leading twist
($t=2$) only of valence $|q\bar{q}\rangle$ and $|gg\rangle$ Fock
states. Let us summarize some important issues:
\begin{itemize}
\item The $\eta - \eta^\prime$ system shows flavor mixing due to the
$SU(3)_{\rm F}$ symmetry breaking and the $U(1)_{\rm A}$ axial
anomaly.
\item The quark-singlet $|\eta_1\rangle$ and the gluonium
state $|gg\rangle$ mix under evolution; both carry flavor-singlet
quantum numbers.
\item The gluon content of the $\eta^\prime$ can reach the level of
$26\%$ \cite{Kou01}.
\item The meson-photon transition form factor contains a singlet
and an octet part: $F_{\rm M\gamma}(Q^2)= F_{\rm M\gamma}^{1}(Q^2)
+ F_{\rm M\gamma}^{8}(Q^2)$.
\item The singlet part, $F_{\rm M\gamma}^{1}(Q^2)$, has a quark and a
gluonic component (\cite{KP03}---using the standard HSA;
\cite{AS03}---using the endpoint-sensitive RC method). This means
\begin{equation}
F_{\rm M\gamma}^{\,1g}(Q^{2})  = \left( T_{\rm
H}^{1}(x,Q^{2},\mu_{\rm F}^2) \quad
 T_{\rm H}^{g}(x,Q^{2},\mu_{\rm F}^2)  \right)
 \, \otimes \,
 \left(
\begin{array}[c]{c}
 \phi_{1}(x,\mu_{\rm F}^2) \\[0.2cm]
 \phi_{g}(x,\mu_{\rm F}^2)
\end{array}
 \right)\, ,
\label{eq:gluon-comp-singlet}
\end{equation}
where $\otimes$ denotes integration over longitudinal momentum
fractions $x$ from $0$ to $1$ and $\mu_{\rm F}$ is the
factorization scale.
\end{itemize}

\begin{figure}[H]
\parbox{0.4\textwidth}
   {\epsfig{file=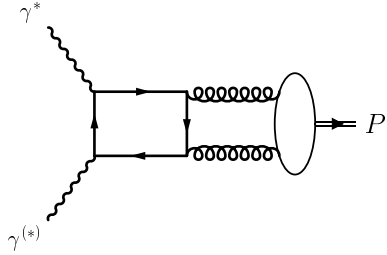,width=0.4\textwidth}}
\hfill
\parbox{0.5\textwidth}
  {\caption{\label{fig:diagram} \small A sample of a Feynman diagram
  contributing to the (pseudoscalar) meson-photon transition form
  factor at NLO.}}
\end{figure}

The transition form factor in the HSA\ can be expressed in terms
of the convolution
\begin{equation}
 F_{\rm M\gamma}(Q^2)
  =
  \int_{0}^{1} dx \phi_{\rm M}(x,\mu_{\rm F}^2)
  T_{\rm H}(x,Q^2;\mu_{\rm F}^2,\mu_{\rm
  R}^2)
\label{eq:trans-FF-conv}
\end{equation}
with $Q^2=-q^2>0$ and $q$ is the four-momentum of the virtual
photon. Figure \ref{fig:diagram} shows an example of the Feynman
diagrams contributing to $F_{\rm M\gamma}$ at NLO.

In Fig.\ \ref{fig:diagram}, the partonic subprocess $\gamma* +
\gamma \to q + \bar{q}$ is described by the hard-scattering
amplitude $T_{\rm H}(x,Q^2;\mu_{\rm F}^2,\mu_{\rm R}^2)$
($\mu_{\rm R}$ being the renormalization scale), whereas the
nonperturbative dynamics is contained in the universal meson
distribution amplitude $\phi_{\rm M}(x,\mu_{\rm F}^2)$. Note that
$\phi_8$ satisfies a scalar evolution equation, analog to the
$\pi$ case, while $\phi_1$ and $\phi_g$ evolve together via a
$(2\!\times\! 2)$-matrix evolution equation. Thus, we have
\cite{ERBL}
\begin{equation}
\frac{d \phi_8(x,\mu_{\rm F}^2)}{d\ln  \mu_{\rm F}^2} =
    V(x,u,\alpha_S(\mu_{\rm F}^2)) \, \otimes \,
\phi_8(u,\mu_{\rm F}^2)
\label{eq:evol-eq}
\end{equation}
with a LO solution given in terms of Gegenbauer polynomials
$(\bar{x}\equiv 1-x)$:
\begin{equation}
\phi_{8} (x,\mu_{\rm F}^2)  =  6 x (1-x) \left[ 1
        +
   \!\!\!\sum_{n=2, 4, \ldots} B_{n}^{\,8}(\mu_{\rm F}^2) \;
          C_{n}^{\,3/2}(2 x -1) \right]
\label{eq:gegenbauer}
\end{equation}
and
\begin{equation*}
\phi_8(x, \mu_{\mathrm{F}}^2)=\phi_8(\bar {x},
\mu_{\mathrm{F}}^2)\, .
\end{equation*}
In Eq.\ (\ref{eq:gegenbauer}), the projection coefficients $B_n^8$
encode the nonperturbative information that is not amenable to QCD
perturbation theory, as we have already mentioned. On the other
hand, the singlet $\phi_1$ and gluonium $\phi_g$ distribution
amplitudes fulfill the matrix evolution equation
\begin{eqnarray}
   \frac{d}{d\ln \mu_{\rm F}^2}
\left(\!\!\!
\begin{array}[c]{c}
\phi_1(x,\mu_{\rm F}^2)\!\!\! \\
\phi_g(x,\mu_{\rm F}^2)\!\!\!
\end{array}
\!\right)= \left(\!\!\!
\begin{array}{cc}
V_{qq} & V_{qg} \\[0.2cm]
V_{gq} & V_{gg}
\end{array}
\!\!\!\right)(x,u,\alpha_S(\mu_{\rm F}^2)) \otimes
 \left(\!\!\!
\begin{array}[c]{c}
\phi_1(u,\mu_{\rm F}^2) \\[0.2cm]
\phi_g(u,\mu_{\rm F}^2)
\end{array}
\!\!\!\right)
\end{eqnarray}
with LO solutions provided by
\begin{eqnarray}
\phi_1(x,\mu_{\rm F}^2) \!\!\!&=&\!\!\! 6 x (1-x) \left[ 1 +
\!\!\!\sum_{n=2,4,\ldots}\!\!\!\!\!\! {\  B_n^{1}(\mu_{\rm F}^2)}
\:  C_n^{3/2}(2 x -1) \right]
 \nonumber \\[0.5cm]
\phi_g(x,\mu_{\rm F}^2) \!\!\!&=&\!\!\!  x (1-x)
 \!\!\!\sum_{n=2,4,\ldots}\!\!\!
 { B_n^{g}(\mu_{\rm F}^2)} \:  C_{n-1}^{5/2}(2 x -1) \;.
\end{eqnarray}
The normalization conditions are
\begin{equation}
\int_0^1 dx \phi_{1,8} (x, \mu_{\rm F}^2) = 1 \qquad \int_0^1 dx
\phi_g (x, \mu_{\rm F}^2) = 0 \, .
\label{eq:normalization}
\end{equation}
The quark component of the singlet state reads
{\footnotesize{
\begin{equation}
\phi_1(x,\mu_{\rm F}^2)  =
 6x\overline{x}\left(
1+\sum_{n=2,4..}^\infty
\left\{ B_n^q\left[ \frac{\alpha_{{\rm s}}(\mu_0^2)}{%
\alpha_{{\rm s}}(\mu_{\rm F}^2)}%
            \right]^{\frac{\gamma_{+}^n}{\beta_0}}
 + \rho_n^gB_n^g\left[ \frac{\alpha_{{\rm
s}}(\mu_0^2)}{\alpha_{{\rm s}}(\mu_{\rm F}^2)}
             \right]^{\frac{\gamma_{-}^n}{\beta_0}}
\right\} C_n^{3/2}(x-\overline{x})\right)
\label{eq:quark-comp-singlet}
\end{equation}}}
with the symmetry condition $\phi_1(x,\mu_{\rm
F}^2)=\phi_1(\bar{x},\mu_{\rm F}^2)$, whereas the gluon component
is {\footnotesize{\begin{equation} \phi_g(x,\mu_{\rm F}^2) =
x\overline{x}\sum_{n=2,4..}^\infty \left\{ \rho_n^qB_n^q\left[
\frac{\alpha_{{\rm s}}(\mu_0^2)}{\alpha_{{\rm s}}(\mu_{\rm F}^2)}
\right]^{\frac{\gamma_{+}^n}{\beta_0}} + B_n^g\left[
\frac{\alpha_{{\rm s}}(\mu_0^2)}{\alpha_{{\rm s}}(\mu_{\rm F}^2)}
\right]^{\frac{\gamma_{-}^n}{\beta_0}}\right\}
C_{n-1}^{5/2}(x-\overline{x})
\label{eq:gluo-comp-singlet}
\end{equation}}}
with the symmetry condition $\phi_g(x,\mu_{\rm F}^2)= -
\phi_g(\bar{x},\mu_{\rm F}^2)$. The associated anomalous
dimensions \cite{AS02} are
\begin{equation}
\gamma _{\pm }^n=\frac 12 \left[ \gamma _{qq}^n+\gamma _{gg}^n\pm
\sqrt{(\gamma _{qq}^n-\gamma _{gg}^n)^2+4\gamma _{qg}^n\gamma
_{gq}^n} \right]\, ,
\label{eq:an-dim}
\end{equation}
\begin{eqnarray}
\rho _n^q=6\frac{\gamma _{+}^n-\gamma _{qq}^n}{\gamma _{gq}^n}
\qquad \rho _n^g=\frac 16\frac{\gamma _{gq}^n}{\gamma
_{-}^n-\gamma _{qq}^n}
\label{eq:rhos}
\end{eqnarray}
with
\begin{eqnarray}
\gamma _{qq}^n & = & C_F\left[ 3+\frac
2{(n+1)(n+2)}-4\sum_{j=1}^{n+1}\frac
1j\right]\, , \nonumber \\[0.5cm]
\gamma _{gg}^n & = & N_c\left[ \frac{\beta _0}{N_c}+\frac
8{(n+1)(n+2)}-4\sum_{j=1}^{n+1}\frac 1j\right] \, ,
\end{eqnarray}
\begin{eqnarray}
\gamma _{qg}^n=\frac{12n_f}{(n+1)(n+2)},\,\,\,\,\,\gamma
_{gq}^n=C_F\frac{n(n+3)}{3(n+1)(n+2)}\, .
\end{eqnarray}
The numerical values of these parameters ($n_f=3$) are
\begin{eqnarray}
\gamma _{qq}^2 & = & -\frac{50}9,\,\,\,\,\,\gamma _{gg}^2=-11,
\,\;\;\;\,\,\gamma_{gq}^2=\,\frac{10}{27},\,\,\,\,\gamma _{qg}^2=3
\nonumber \\
\gamma _{+}^2 & \simeq & -\frac{48}9,\,\,\,\gamma _{-}^2 \simeq
-\frac{101}9\,,\,\,\,\rho_2^q \simeq \frac{16}5,\,\,\,\,\,\rho
_2^g\simeq -\frac1{90}\, . \label{eq:anom-numer-values}
\end{eqnarray}
The required Gegenbauer polynomials are
\begin{eqnarray}
C_2^{3/2}(x-\overline{x}) & = & \frac 32\left[
5(x-\overline{x})^2-1\right]
=6\left( 1-5x\overline{x}\right) \nonumber \\
C_1^{5/2}(x-\overline{x}) & = & 5(x-\overline{x}) \, .
\label{eq:Gegenbauers}
\end{eqnarray}

\section{Hard-scattering amplitudes for the $\eta\gamma$ and
$\eta^{\prime}\gamma$ transition}
\label{sec:HSA-trans}

The form factor for the $\eta\gamma$ and $\eta^{\prime}\gamma$
transition, given by $F_{\rm M\gamma}(Q^2)= F_{\rm
M\gamma}^{1}(Q^2) + F_{\rm M\gamma}^{8}(Q^2)$, contains a singlet
part comprising quark and gluon components:
{\footnotesize{\begin{eqnarray} Q^2F_{\rm M\gamma}^1(Q^2) &
\!\!\!= \!\!\!& \int_0^1 dx f_{\rm M}^1N_1\Bigl\{ T_{\rm H,
LO}^q(x) \phi_{1}(x,\mu_{\rm F}^2) +\; \int_0^1 dx
\frac{\alpha_{{\rm s}}(\mu_{\rm R}^2)}{4\pi} C_{\rm
F} \nonumber\\[0.3cm]
\!\!\!&&\!\!\! \left[ T_{\rm H, NLO}^q(x,Q^2,\mu_{\rm F}^2)
\phi_{1}(x,\mu_{\rm F}^2) +  T_{\rm H,NLO}^g(x,Q^2,\mu_{\rm F}^2)
\phi_{g}(x,\mu_{\rm F}^2)\right] \Bigr\}\, .
\label{eq:FF-singlet-quark-gluon}
\end{eqnarray}}}
The octet part contains only a quark component; it reads
\begin{eqnarray} Q^2F_{\rm M\gamma }^8(Q^2) & = & \int_0^1 dx
f_{\rm M}^8N_8\left[ \!\!\!\!\!\!\!
\!\!\!\!\!\!\!\!\!\!\phantom{\frac{\alpha_{{\rm s}}(\mu_{\rm
R}^2)}{4\pi}}T_{\rm H, LO}^q(x) \phi_{8}(x,\mu_{\rm
F}^2)\right. \nonumber \\[0.2cm]
&&  \left. + \; \frac{\alpha_{{\rm s}}(\mu_{\rm R}^2)}{4\pi} \,
C_{\rm F} T_{\rm H, NLO}^q(x,Q^2,\mu_{\rm F}^2)
\phi_{8}(x,\mu_{\rm F}^2)\;\right] \, .
\label{eq:FF-octet-quark}
\end{eqnarray}
The expressions for the involved hard-scattering amplitudes are
\begin{equation}
T_{\rm H, LO}^q(x)=x^{-1}+\bar{x}^{-1} \label{eq:HSA-quark-LO}\, ;
\end{equation}
\begin{equation} \!\!\!\!\!T_{\rm H, NLO}^q(x,Q^2,\mu_{\rm F}^2)
=\frac 1x\left[ \ln {}^2x-\frac{x\ln x}{\overline{x}} -9\right]
+\frac 1x(3+2\ln x)\ln \frac{Q^2}{\mu_{\rm F}^2}+(x\leftrightarrow
\overline{x})
\label{eq:HSA-quark-NLO}
\end{equation}
\begin{eqnarray} \!\!\!\!\!T_{\rm H,
NLO}^g(x,Q^2,\mu_{\rm F}^2) =\frac{x\ln
{}^2x}{\overline{x}}+\left( 6-\frac 4{\overline{x}}\right) \ln
{}x+2\frac{x\ln x}{\overline{x}}\ln \frac{Q^2}{\mu_{\rm F}^2}
-(x\leftrightarrow \overline{x})
\label{eq:HSA-gluon-NLO}
\end{eqnarray}
and the charge factors read
\begin{equation}
N_1=\frac 1{\sqrt{3}}\left( e_u^2+e_d^2+e_s^2\right) ,\;\;\;\;\;
N_8=\frac 1{\sqrt{6}}\left( e_u^2+e_d^2-2e_s^2\right) \, .
\label{eq:chrages}
\end{equation}

\section{$\eta\gamma$,\;$\eta^\prime\gamma$ transition form factor in the
         RC\ approach}
\label{sec:trans-RC-appr}

Let us outline here the essentials of the endpoint-sensitive RC\
method.
\begin{itemize}
\item Solve the renormalization group equation for
$\alpha_s(\lambda Q^2)$ in terms of $\alpha_s(Q^2)$ \cite{CS94} to
$\alpha_s^2(Q^2)$ accuracy.
\item Expand the hard-scattering amplitude $T(Q^2)$ of the process
as a power series in $\alpha_s(Q^2)$ with factorially growing
coefficients $C_n \sim (n-1)!$.
\item Use the Borel integral technique to resum them by
\begin{itemize}
\item determining first the Borel transform ${\cal B}[T]$ of this
series
\item inverting then ${\cal B}[T]$ to get
$$\!\!\!\!\!\!\!\!\!\!\![T]^{\rm resum}(Q^2)
\sim \mbox{P.V.} \int_{0}^{\infty} du
\exp\left[\frac{-4\pi u}{\beta_0\alpha_s(Q^2)}\right]
{\cal B}[T](u).$$ \\[-0.5cm]
\end{itemize}
At this point a couple of important remarks are in order. (i) The
Borel transforms contain poles on the positive $u$ axis that are
exactly \emph{IR renormalon poles}; hence a principal value (P.V.)
prescription has to be used. (ii) A direct way to obtain the Borel
resummed expressions is via the \emph{Inverse Laplace
Transformation}.

Then, one finds
\begin{equation}
\alpha_s(xQ^2) =
  \frac{4\pi}{\beta_0}\int_0^\infty du {\rm e}^{-ut} R(u,t) x^{-u}
\label{eq:alpha-s}
\end{equation}
with
\begin{equation}
  R(u,t)
=
  1- \frac{2\beta_1}{\beta_0^2} u(1-\gamma_{\rm E} -\ln t -\ln u)
  \, .
\label{eq:Inv-Lapl-trans}
\end{equation}
\item Endpoint singularities $x\to 1\; \bar{x}\to 1$
transform into IR renormalon (multi-)pole divergences at $u_{0}=n$
(in our case $n=1,2,3,4$) in the Borel $u$ plane.
\item Removing these poles via the P.V. prescription, we obtain
\emph{resummed} expressions for
$$
  [Q^2F_{{\rm M}\gamma}^1(Q^2)]^{\rm resum} \; ,\qquad
  [Q^2F_{{\rm M}\gamma}^8(Q^2)]^{\rm resum}\, .
$$
\item The pole at $u_0=n$ of the Borel plane corresponds to
power-suppressed corrections $\sim (1/Q^2)^n$ contained in the
scaled form factors.
\end{itemize}

Let us close this section, by commenting upon the importance of
power corrections from the theoretical point of view and in
comparison with the standard HSA. The latter prefers to set
$\mu_{\rm R}^2=Q^2$. Then, large NLO\ logarithms are present. The
RC\ method sets instead $\mu_{\rm R}^2=xQ^2$. As a result, the
term $\ln (\mu_{\rm R}^2/xQ^2)$ in the NLO contribution is
eliminated, but the integration over $x$ gives rise to
\emph{power-suppressed contributions} in the endpoint regions $x
\to 0,1$. Note in this context that because asymptotically both
approaches have to yield the same results, one has to verify that
the induced power corrections do not affect this regime, leaving
the asymptotic behavior of perturbative QCD unchanged. Hence, in
technical terms, one has to ensure that
$$
 \int_0^\infty du {\rm e}^{-ut} R(u,t)
\stackrel{Q^2\to\infty}{\to}
 \int_0^\infty du {\rm e}^{-ut} \, .
$$
In view of the above remarks, the best (perturbative) procedure is
the one that minimizes the NLO contribution while keeping power
corrections under control.

The present analysis employs the following scales:
\begin{itemize}
\item $\mu_{\rm R}^2=xQ^2$ (renormalization scale)
\item $\Lambda^{(n_f=4)}=0.25$~GeV
\item $\mu_0^2=1$~GeV$^2$ (normalization scale)
\item $\mu_{\rm F}^2=Q^2$ (factorization scale)
\end{itemize}
The estimated influence of higher-twist uncertainties is of the
order of $(10-15)\%$.

\section{Borel Resummed $\eta\gamma$ and $\eta^\prime\gamma$
transition form factors}
\label{sec:Borel-res-FF}

The NLO expression for the transition form factor, calculated with
the RC method \cite{AS03}, comprises a quark component
\begin{eqnarray}
  Q^2F_{\rm M\gamma}^1(Q^2)_1^{\rm quark}
& \sim &
    \alpha_{{\rm s}}(Q^2x)t(x,\mu_{\rm F}^2)\otimes
    \phi_{1}(x,\mu_{\rm F}^2)\nonumber \\
& + &
    \alpha_{{\rm s}}(Q^2\overline{x})t(\overline{x},\mu_{\rm F}^2)
\otimes
    \phi_{1}(x,\mu_{\rm F}^2)\nonumber\\
& = &
    2\alpha_{{\rm s}}(Q^2x)t(x,\mu_{\rm F}^2)
    \otimes
    \phi_{1}(x,\mu_{\rm F}^2)
\label{eq:FF-quark-comp}
\end{eqnarray}
with
\begin{equation}
t(x, \mu_{\rm F}^2)=\frac 1x\left[ \ln {}^2x-\frac{x\ln
x}{\bar{x}} -9\right] +\frac 1x(3+2\ln x)\ln \frac{Q^2}{\mu_{\rm
F}^2} \label{eq:hard-part}
\end{equation}
and a gluon component
\begin{equation}
Q^2F_{\rm M\gamma}^1(Q^2)_1^{\rm gluon} \sim
  2\alpha_{{\rm s}}(Q^2x)g(x,\mu_{\rm F}^2)
\otimes
  \phi_g(x,\mu_{\rm F}^2)
\label{eq:FF-gluon-comp}
\end{equation}
with
\begin{equation}
g(x,\mu_{\rm F}^2)=\frac{x\ln {}^2x}{\overline{x}}
 +\left( 6-\frac 4{\overline{x}}\right)
 \ln x +2\ln\left(\frac{Q^2}{\mu_{\rm
 F}^2}\right)
 \frac{x \ln x}{\overline{x}} \, .
\label{eq:gluon-singlet-DA}
\end{equation}
Summing up, we can write--in the context of the RC\ method--the
transition form factors $Q^2F_{\rm M\gamma}^1(Q^2)$ and $Q^2F_{\rm
M\gamma}^8(Q^2)$ as follows
\begin{eqnarray}
  Q^2F_{\rm M\gamma}^1(Q^2)
\!\!\!& = &\!\!\!
  f_{\rm M}^1N_1\left\{ T_{{\rm H},0}^q(x)
\otimes
  \phi_{1}(x,\mu_{\rm F}^2)
+  \frac{C_{\rm F}}{2\pi }
   \left[ \alpha _{{\rm s}}(Q^2x)t(x,\mu_{\rm F}^2)\right.\right.
\nonumber\\[0.3cm]
&&\!\!\!\!\! \otimes  \!\left.\left.
  \phi_{1}(x,\mu_{\rm F}^2)
+
 \alpha _{{\rm s}}(Q^2x)g(x,\mu_{\rm F}^2)
\otimes
  \phi_{g}(x,\mu
_F^2)\right]
\phantom{\frac{\alpha_H}{2\pi_H}}\!\!\!\!\!\!\!\!\!\!\!\!\!\right\}
\label{eq:singlet-FF-RC-meth}
\end{eqnarray}
and
\begin{eqnarray}
  Q^2F_{\rm M\gamma }^8(Q^2)
\!\!&=&\!\!
  f_{\rm M}^8N_8 \left[ T_{{\rm H},0}^q(x)
\otimes \phi_{8}(x,\mu_{\rm F}^2) + \frac{C_{\rm
F}}{2\pi}\,\alpha_{{\rm s}}(Q^2x)t(x,\mu_{\rm F}^2)\right.
\nonumber \\[0.3cm]
&& \otimes \left.
  \phi_{8}(x,\mu_{\rm F}^2)
  \phantom{\frac{\alpha_H}{2\pi_H}}\!\!\!\!\!\!\!\!\!\!\!\!\!\right]\,
  .
\label{eq:octet-FF-RC-meth}
\end{eqnarray}
Now recall that the running coupling $\alpha_{s}(xQ^2)$ in terms
of $\alpha_{s}(Q^2)$ \cite{CS94} reads
$$
  \alpha_{\rm s}(Q^2x) \simeq \frac{\alpha_{\rm s}(Q^2)}{1+\ln x/t}
  \left[1 - \frac{\alpha_{\rm s}(Q^2)\beta_1}{2\pi\beta_0}
  \frac{\ln [1+ \ln x/t]}{1+ \ln x /t}\right]
$$
where we have used
$$
 t=\frac{4\pi}{\beta_0\alpha_{\rm s}(Q^2)}=\ln
 \frac{Q^2}{\Lambda^2}\, ,
\quad \beta_0=11-\frac2{3} n_f\, , \quad
\beta_1=51-\frac{19}{3}n_f \, .
$$

In this way, we finally arrive at the following expressions for
the resummed singlet and octet transition form factors within the
RC method:
\begin{eqnarray}
  Q^2F_{\rm M\gamma}^1(Q^2)
\!\!\!& = &\!\!\!
    f_{\rm M}^1N_1\left\{ 6+A(\mu_{\rm F}^2)
  + \frac{12C_{\rm F}}{\beta_0}
  \left[\!\!\!\!\!\!\!\!\!\!\!\!\phantom{\frac{\alpha_H}{2\pi_H}}
  \left( 1+A(\mu_{\rm
  F}^2)\right)\nonumber \right.\right.\\[0.3cm]
 &\times & \!\!\!\left. \left. \int_0^\infty \!\!\! du
 {\rm e}^{-ut}R(u,t)Q_1(u)
-  5A(\mu_{\rm F}^2)\int_0^\infty
   \!\! du {\rm e}^{-ut}R(u,t)Q_2(u)\right]\nonumber\right. \\[0.3cm]
 & +&\!\!\!\left. \frac{2C_{\rm F}}{\beta_0}B(\mu_{\rm F}^2)
 \int_0^\infty \!\!\! du {\rm e}^{-ut}R(u,t)G(u)\right\} \; ,
\label{eq:res-FF-singlet}
\end{eqnarray}
\begin{eqnarray}\!\!\!\!\!\!\!\!
  Q^2F_{\rm M\gamma }^8(Q^2)
& = &
    f_{\rm M}^8N_8\left\{ 6+C(\mu_{\rm F}^2)
  + \frac{12C_{\rm F}}{\beta _0}\left[ \left( 1+C(\mu_{\rm F}^2)\right)
    \int_0^\infty du {\rm e}^{-ut}\right.\right. \nonumber \\[0.5cm]
\;\;\; & \times & \!\!\!\left.\left.R(u,t)Q_1(u)-\right. \right.
    \left. \left. 5C(\mu_{\rm F}^2)\int_0^\infty du
    {\rm e}^{-ut}R(u,t)Q_2(u)\right]
    \right\}\, .
\label{eq:res-FF-octet}
\end{eqnarray}
In the above expressions, the following abbreviations have been
used \cite{AS03}:
\begin{equation}
 R(u,t)=1-\frac{2\beta_1}{\beta_0^2}u(1-\gamma_E-\ln t-\ln u)\, ,
\label{eq:R}
\end{equation}

$$
Q_1(u)=\frac 2{(1-u)^3}-\frac 2{(2-u)^3}-\frac{2a}{(1-u)^2}+\frac
{1+2a}{(2-u)^2}+3\frac{a-3}{(1-u)(2-u)}\, ,
$$

\begin{eqnarray}
Q_2(u)&=&\frac 2{(2-u)^3}-\frac 4{(3-u)^3}+ \frac
2{(4-u)^3}-\frac{2a}{(2-u)^2}\nonumber \\[0.3cm]
 &\,+\,& \frac {1+4a}{(3-u)^2}-\frac
{1+2a}{(4-u)^2}+6\frac{a-3}{(2-u)(3-u)(4-u)}\, ,
\end{eqnarray}

\begin{equation}
G(u)=\frac 4{(4-u)^3}-\frac 2{(3-u)^3}+\frac 2{(2-u)^2}
-2\frac{5-a}{(3-u)^2}+4\frac{3-a}{(4-u)^2}\, ,
\label{eq:Qs-G}
\end{equation}
with $a\equiv \ln(Q^2/\mu_{\rm F}^2)$ and the Gegenbauer
coefficients being given by
\begin{eqnarray}
A(\mu_{\rm F}^2) & = & 6B_2^q\left[ \frac{\alpha_{{\rm
s}}(\mu_{\rm F}^2)}{\alpha_{{\rm s}}(\mu_0^2)}
\right]^{\frac{48}{81}}-\frac{B_2^g}{15} \left[\frac{\alpha _{{\rm
s}}(\mu_{\rm F}^2) }{\alpha_{{\rm s}}(\mu_0^2)}
\right]^{\frac{101}{81}} \nonumber \\[0.5cm]
B(\mu_{\rm F}^2) & = &
16B_2^q\left[ \frac{\alpha_{{\rm s}}(\mu_{\rm F}^2)}{\alpha_{%
{\rm s}}(\mu_0^2)}\right]^{\frac{48}{81}}
+5B_2^g\left[\frac{\alpha_{{\rm s}}(\mu_{\rm F}^2)}{\alpha_{{\rm
s}}(\mu_0^2)} \right]^{\frac{101}{81}} \nonumber \\[0.5cm]
C(\mu_{\rm F}^2) & = &
6B_2^q\left[ \frac{\alpha _{{\rm s}}(\mu_{\rm F}^2)}{\alpha _{{\rm %
s}}(\mu _0^2)}\right] ^{\frac{50}{81}} \, .
\label{eq:gegen}
\end{eqnarray}

\section{Phenomenological Analysis}
\label{sec:pheno-ana}
\begin{figure}[]
\parbox{0.45\textwidth}
   {\epsfig{file=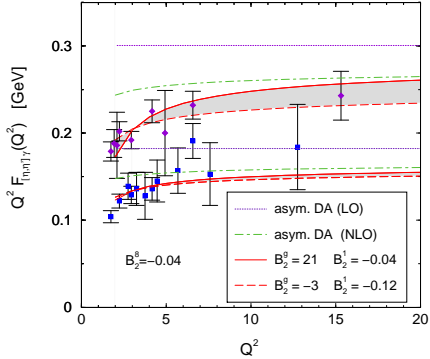,width=0.45\textwidth}}
\hfill
\parbox{0.5\textwidth}
  {\caption{\label{fig:Kroll-Passek} \small $\eta$ (lower curves) and
    $\eta^\prime$ (upper curves) transition form factors obtained with
    the standard HSA in \protect\cite{KP03}. The shaded area
    corresponds to the range of values $B_2^8(\mu_0^2)=-0.04\pm 0.04
    \quad B_2^1(\mu_0^2)=-0.08\pm 0.04 \quad B_2^g(\mu_0^2)=9\pm 12$
    (see \protect\cite{KP03,Pas03} for further details).}}
\end{figure}

In this section we perform numerical computations of the Borel
resummed and rescaled by $Q^2$ $\eta \gamma$ and
$\eta^{\prime}\gamma$ transition form factors in order to extract
the $\eta$ and $\eta^{\prime}$ meson distribution amplitudes from
the CLEO\ data. We shall also compare our theoretical predictions
with those obtained with the standard HSA\ \cite{KP03,AP03}, the
aim being to reveal the role of power corrections at low-momentum
transfer in the exclusive process under consideration.

Let us start our discussion by quoting the results obtained in
\cite{KP03} (see also \cite{Pas03}) using the standard HSA. Their
main predictions are shown in Fig.\ \ref{fig:Kroll-Passek} in
comparison with the CLEO data \cite{CLEO98}.

\begin{figure}[t]
\epsfig{file=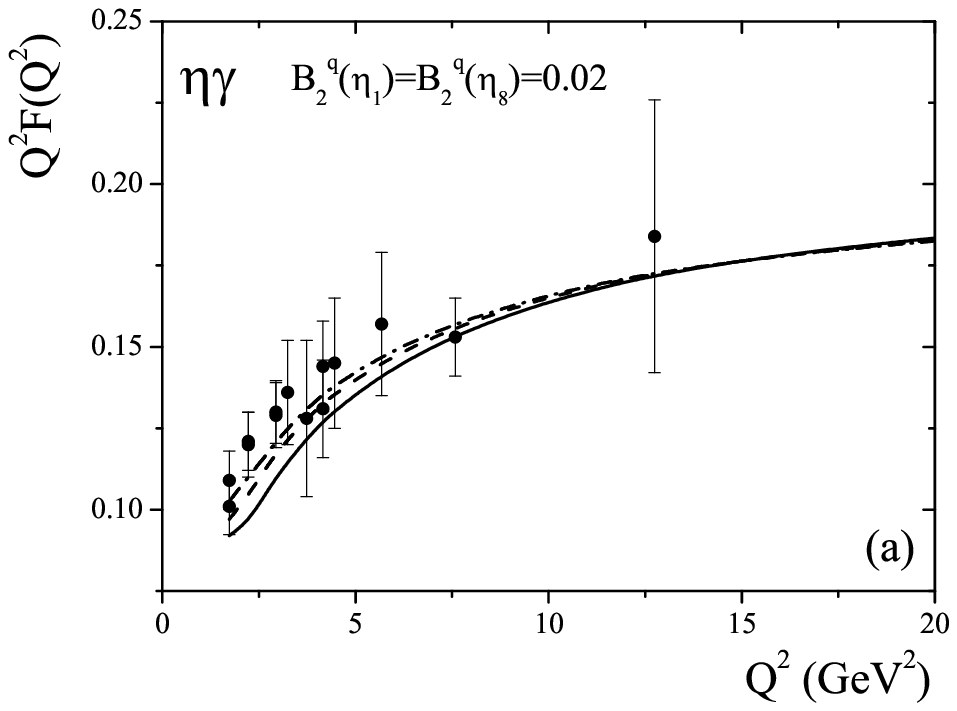,width=0.45\textwidth}
\epsfig{file=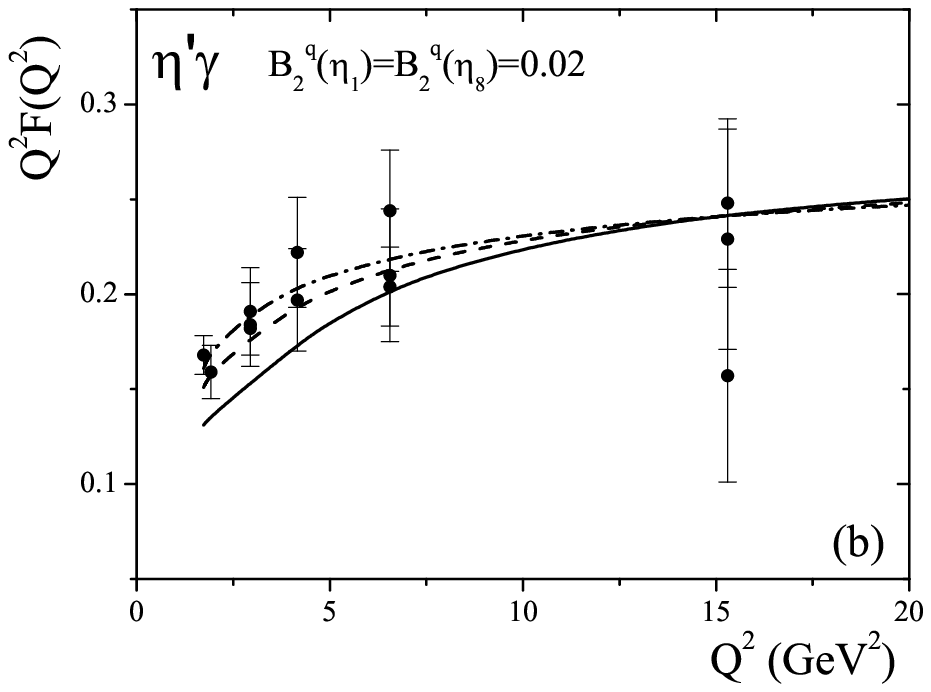,width=0.45\textwidth}
    \caption{\label{fig:AS-fig1} \small Predictions for the scaled
form factors as functions of $Q^2$ of the $\eta\gamma$ (left
panel) and $\eta^{\prime}\gamma$ (right panel) electromagnetic
transition. For the solid curves the designation is $B_2^g(\eta
_1)=0$. The dashed lines correspond to $B_2^g(\eta _1)=10$; for
the dash-dotted curves we use $B_2^g(\eta _1)=15$. The data are
taken from Ref.\ \cite{CLEO98}.}
\end{figure}

As one sees from this figure, the agreement between the
theoretical predictions and the low-momentum data is rather
poor---especially when using asymptotic profiles for the $\eta$,
$\eta^\prime$ meson distribution amplitudes. To decrease the
magnitude of the form factors at low $Q^2$, and achieve this way a
better agreement with the data, the standard HSA\ would call for
the two-angles mixing scheme and for distribution amplitudes
mainly with $B_2^q(\eta_1),\,B_2^q(\eta_8)<0$. The inclusion of
power-law corrections changes the low-momentum behavior of the
form-factor predictions significantly, as one observes from Fig.\
\ref{fig:AS-fig1}. Indeed, using the standard octet-singlet mixing
scheme, one can reproduce the trend of the CLEO data rather well
in the whole momentum range explored---especially with a non
negligible gluon contribution (the Gegenbauer coefficients are
given in Fig.\ \ref{fig:AS-fig1})---because the effect of power
corrections is to enhance the absolute value of the NLO correction
to the form factors by more than a factor of $2.5-3$. Since the
contribution of the NLO term to the form factors is negative, the
power corrections reduce the leading-order prediction for the form
factors considerably, while at the highest $Q^2$ values measured
by the CLEO collaboration this influence becomes more moderate.

\begin{figure}[H]
\epsfig{file=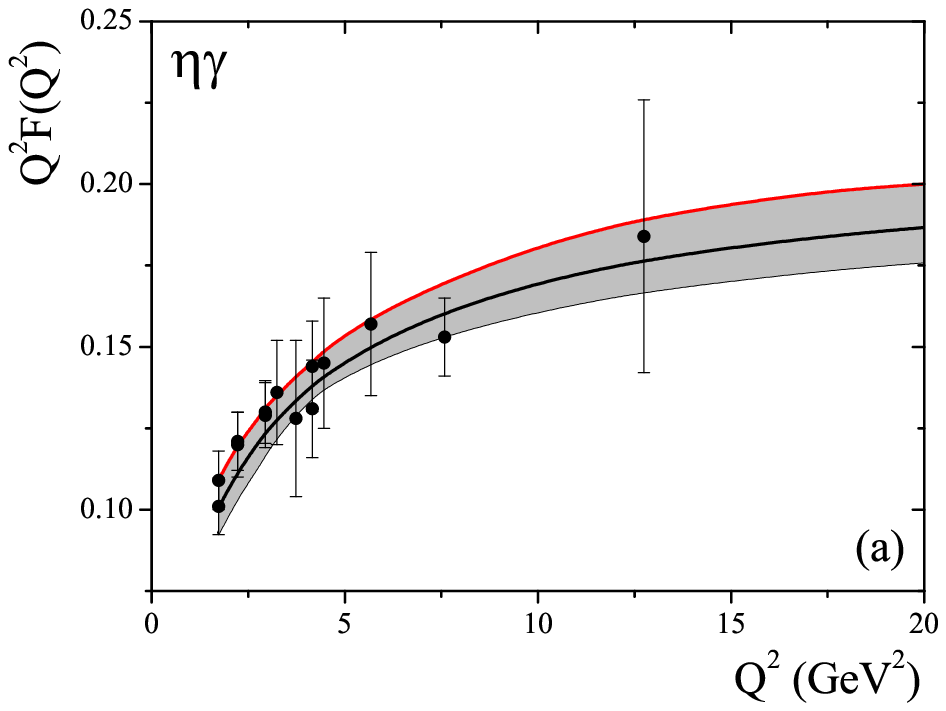,width=0.45\textwidth}
\epsfig{file=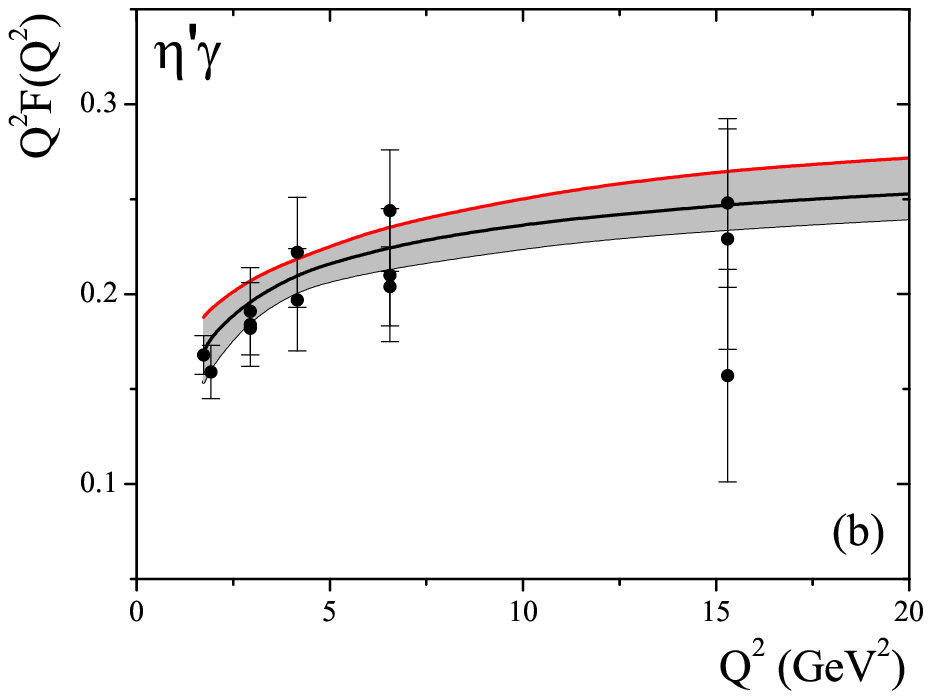,width=0.45\textwidth}
    \caption{\label{fig:AS-fig2} \small The $\eta\gamma$ (left) and
$\eta^{\prime}\gamma$ (right) scaled transition form factors as
functions of $Q^2$. The central solid curves are found using the
values $B_2^q(\eta_1)=B_2^q(\eta_8)=0.05$ and $B_2^g(\eta_1)=17$.
The shaded areas demonstrate the $1\sigma$ regions for the
transition form factors.}
\end{figure}

The $1\sigma$ regions in the form of shaded areas for the scaled
form factors for the $\eta\gamma$ and $\eta^\prime\gamma$
transition in the RC method and using the octet-singlet scheme are
displayed in Fig.\ \ref{fig:AS-fig2}. The central line corresponds
to the coefficients values $B_2^q(\eta_1)=B_2^q(\eta_8)=0.05$;
\quad $B_2^g(\eta_1)=17$. A full-fledged discussion of these
issues is given in \cite{AS03}, together with error estimates
arising from varying the values of the theoretical parameters used
in the analysis.

\begin{figure}[H]
\epsfig{file=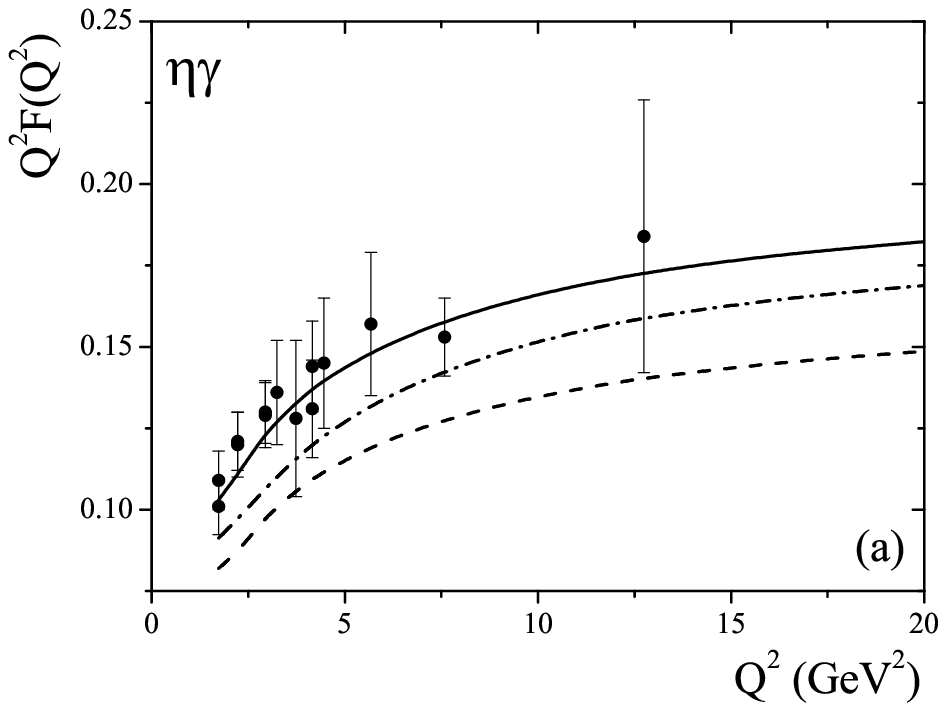,width=0.45\textwidth}
\epsfig{file=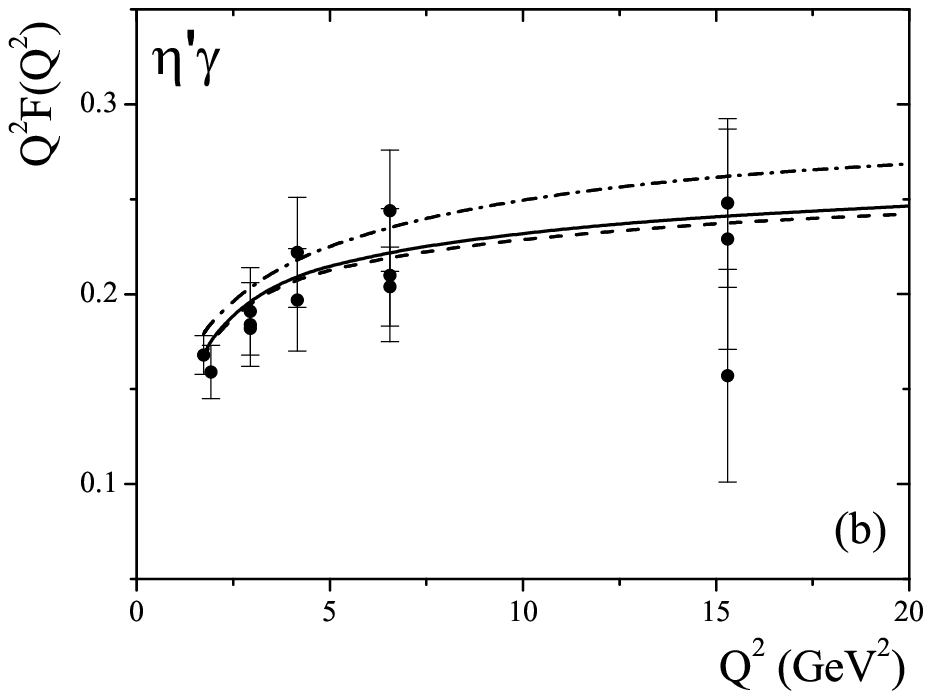,width=0.45\textwidth}
    \caption{\label{fig:AS-fig3} \small The $\eta\gamma$ (a) and
$\eta^{\prime}\gamma$ (b) electromagnetic transition form factors
vs.\ $Q^2$. The solid lines correspond to the ordinary
octet-singlet mixing scheme with parameters
$B_2^q(\eta_1)=B_2^q(\eta_8)=0.02$ and $B_2^g(\eta_1)=18$. The
broken lines are obtained within the two-angles mixing scheme. The
dashed lines describe the situation with the same parameters as
the solid curves. The parameters for the dash-dotted curves are
$B_2^q(\eta_1)=B_2^q(\eta_8) =0.15,\,B_2^g(\eta_1)=18$.}
\end{figure}

It is important to emphasize that our calculations do not exclude
the usage of the two-angles mixing scheme in conjunction with the
RC\ method. But in such a case, a considerably larger contribution
of the non-asymptotic terms to the distribution amplitudes of the
$\eta_1$ and $\eta_8$ states would be required. Carrying out such
a computation \cite{AS03}, we obtained the results shown in Fig.\
\ref{fig:AS-fig3}. Inspection of the left panel of this figure
reveals that the $\eta\gamma$ transition FF\ found within this
scheme lies significantly lower than the data. Therefore, to
improve the agreement with the experimental data, a relatively
large contribution of the first Gegenbauer polynomial to the
distribution amplitudes of the $\eta_1$ and $\eta_8$ states seems
necessary. The Gegenbauer coefficients corresponding to the
predictions shown in Fig.\ \ref{fig:AS-fig3} are
$B_2^q(\eta_1)=0.15,\;B_2^q(\eta_8)=0.15$ and $B_2^g(\eta_1)=18$.
We consider the values $B_2^q(\eta_1)=B_2^q(\eta_8)=0.15$ as
actually determining the \emph{lower} bound for the admissible set
of distribution amplitudes in the context of the two-angles mixing
parameterization scheme. Hence, in that scheme, we obtain
\begin{equation}
 B_2^q(\eta_1)=B_2^q(\eta_8)=0.15,\;B_2^g(\eta_1) \in [16,\,20]\, .
\label{eq:gegen-2mix-angles}
\end{equation}

\begin{figure}[H]
\parbox{0.45\textwidth}
   {\epsfig{file=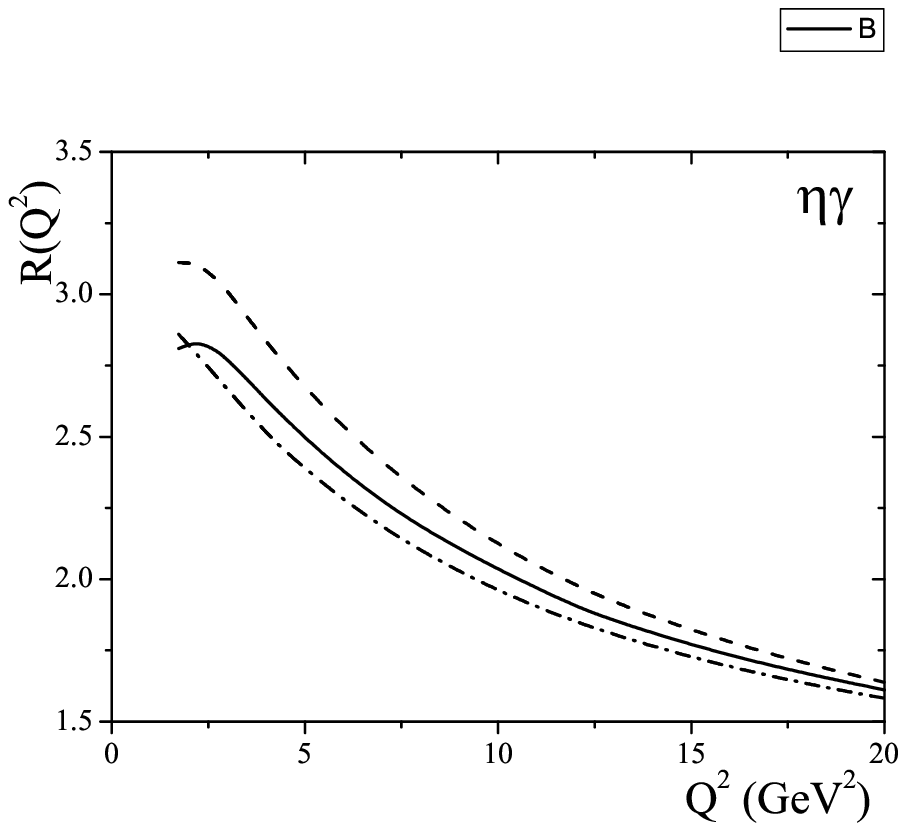,width=0.45\textwidth,clip=}}
\hfill
\parbox{0.5\textwidth}
  {\caption{\label{fig:AS-fig4} \small The ratio $R(Q^2)$ for the
$\eta \gamma$ form factor. The solid line corresponds to the input
parameters $B_2^q(\eta _1)=B_2^q(\eta _8)=B_2^g(\eta _1)=0$. The
dash-dotted curve describes the same ratio, but for $B_2^q(\eta
_1)=B_2^q(\eta _8)=0,\,B_2^g(\eta _1)=14$, while the dashed one
corresponds to $B_2^q(\eta _1)=B_2^q(\eta
_8)=0.05,\,B_2^g(\eta_1)=10$.}}
\end{figure}

Let us close this section by summarizing the main differences
between the standard HSA\ and the RC\ method:  (i) Form factors in
the HSA\ overshoot the CLEO data---especially in the low $Q^2$
region---even with the NLO corrections included. (ii) Values of
the Gegenbauer coefficients $B_2^q(\eta_1), \; B_2^q(\eta_8)>0$
increase the disagreement, while $B_2^g(\eta_1)>0$ reduces the
disagreement. Hence, a better agreement with the CLEO data would
call for the two-angles mixing scheme and $B_2^q(\eta_1)\, ,
B_2^q(\eta_8)<0$. (iii) The inclusion of power corrections
enhances the (negative) NLO correction to the form factors at low
$Q^2$ by factors $2.5 - 3$. In order to quantify these statements,
we show in Fig.\ \ref{fig:AS-fig4}, the numerical results for the
ratio
\begin{equation}
\label{eq:62} R_{\rm M\gamma}(Q^2)
=\frac{[Q^2F_{\rm M\gamma}(Q^2)]_{\rm NLO}^{\rm res}}{%
[Q^2F_{\rm M\gamma }(Q^2)]_{\rm NLO}^{\rm HSA}}
\end{equation}
\label{eq:ratio}
for some selected values of the expansion coefficients. As a
result, the RC\ method, employing the one-angle mixing scheme, is
in good agreement with the CLEO data. (iv) Using instead the
two-angles mixing scheme, the RC\ method favors non-asymptotic
profiles for the distribution amplitudes of $\eta_1$ and $\eta_8$,
e.g., $B_2^q(\eta_1), \; B_2^q(\eta_8)\geq 0.15$ and
$B_2^g(\eta_1)\in [16, \; 20]$, while the region $B_2^q(Q^2\simeq
2~{\rm GeV}^2)<0$ seems to be incompatible with the CLEO data.

\section{Conclusions}
\label{sec:concl}

The renormalon-inspired RC\ method enables the inclusion of power
corrections originating from the kinematic endpoint region ($x\to
0,1$), where nonperturbative QCD dominates and fixed-order
perturbative computations of such corrections yields divergent
results. We found that power-suppressed ambiguities to form
factors vary between $3\%$ at high and $11\%$ at low $Q^2$ values.
On the other hand, we have verified that the asymptotic limit of
$[Q^2F_{{\rm M}\gamma}(Q^2)]^{\rm resum}$ coincides, as it should,
with the standard HSA\ result, leaving the asymptotic properties
of QCD perturbation theory unchanged. The effect of power
corrections at $Q^2\leq 5$~GeV$^2$ enhances the (negative) NLO
correction by $2.5 - 3$ times, providing  this way agreement with
the trend of the CLEO data. In the standard octet-singlet scheme
we found $B_2^q(\eta_1)=B_2^q(\eta_8)\geq 0.055\pm 0.065$, \quad
$B_2^g(\eta_1)=18\pm 4.5$, whereas in the two-angles mixing
scheme, we found $B_2^q(\eta_1)=B_2^q(\eta_8)\geq 0.15$, \quad
$B_2^g(\eta_1) \in [16, \, 20]$. The distribution amplitude of the
$\eta$ and $\eta^{\prime}$ mesons, obtained in this work, can be
useful in the investigation of other exclusive processes that
involve $\eta$ and $\eta^{\prime}$ mesons, especially at lower
momentum-transfer values, where the standard HSA\ is most
unreliable.

\vspace{0.5cm} \centerline{\bf Acknowledgments}\vspace{0.2cm}

One of us (N.G.S.) would like to thank the organizers of the
workshop for the hospitality and the exciting atmosphere during
the meeting.

\end{document}